\documentclass[prl,superscriptaddress,twocolumn]{revtex4-1}
\usepackage{epsfig}
\usepackage{graphicx}% Include figure files
\usepackage{dcolumn}% Align table columns on decimal point
\usepackage{bm}% bold math
\usepackage{tabularx}% bold math
\usepackage{hyperref}
\hypersetup{citecolor=black, filecolor= black, linkcolor= black, urlcolor= black}
\begin{document}

\title{Stability of xenon oxides at high pressures}

\author{Qiang Zhu}
\email{qiang.zhu@stonybrook.edu}
\affiliation{Department of Geosciences, Department of Physics and Astronomy, and New York Center for Computational Sciences, Stony Brook University, Stony Brook, New York 11794, USA}
\author{Daniel Y. Jung}
\affiliation{Laboratory of Crystallography, Department of Materials, ETH Honggerberg, Wolfgang-Pauli-Strasse 10, CH-8093 Z\"urich, Switzerland}
\author{Artem R. Oganov}
\email{artem.oganov@stonybrook.edu}
\affiliation{Department of Geosciences, Department of Physics and Astronomy, and New York Center for Computational Sciences, Stony Brook University, Stony Brook, New York 11794, USA}
\affiliation{Geology Department, Moscow State University, 119992, Moscow, Russia}
\author{Colin W. Glass}
\affiliation{High Performance Computing Center Stuttgart (HLRS), Germany}
\author{Carlo Gatti}
\affiliation{Istituto di Scienze e Tecnologie Molecolari del CNR (CNR-ISTM) e Dipartimento di Chimica Fisica ed Elettrochimica, Universita di Milano, via Golgi 19, I-20133, Milano, Italy}
\author{Andriy O. Lyakhov}
\affiliation{Department of Geosciences, Department of Physics and Astronomy, and New York Center for Computational Sciences, Stony Brook University, Stony Brook, New York 11794, USA}

\begin{abstract}
Xenon, which is quite inert under ambient conditions, may become reactive under pressure. The possibility of formation of stable xenon oxides and silicates in the interior of the Earth could explain the atomspheric missing xenon paradox. Using the \emph{ab initio} evolutionary algorithm, we predict the thermodynamical stabilization of Xe-O compounds at high pressures (XeO, XeO$_2$ and XeO$_3$ at pressures above 83, 102 and 114 GPa, respectively). Our calculations indicate large charge transfer in these oxides, suggesting that large electronegativity difference and pressure are the key factors favoring the formation of xenon compounds. Xenon compounds in the Earth's mantle, however, cannot directly explain the missing xenon paradox: xenon oxides are unstable in equilibrium with metallic iron in the Earth's lower mantle, while xenon silicates are predicted to spontaneously decompose at all mantle pressures ($<$136 GPa). This does not preclude Xe atoms from being retained in defects of mantle silicates and oxides.
\end{abstract}

\maketitle

Xenon is a noble gas, chemically inert at ambient conditions. A few xenon fluorides have been found \cite{Levy-JACS-1963, Templeton-JACS-1963, Hoyer-JFC-2006}, with Xe atoms in the oxidation states +2, +4, +6, or +8. Upon application of high pressure, the molecular phase of insulating XeF$_2$ has been reported to transform into two- and three-dimensional extended solids and become metallic \cite{Kim-NC-2010}. Clathrate Xe-H solids were also observed at pressures above 4.8 GPa \cite{Somayazulu-NC-2010}. Two xenon oxides (XeO$_3$, XeO$_4$) \cite{Smith-JACS-1963} are known at atmospheric pressure, but are unstable and decompose explosively above 25$^\circ$ C (XeO$_3$) and -40$^\circ$ C (XeO$_4$) \cite{Selig-Science-1964}. A crystalline XeO$_2$ phase with local square-planar XeO$_4$ geometry has recently been synthsized at ambient conditions \cite{Brock-JACS-2011}.

There is growing evidence to suggest that noble gases, especially Xe, may become reactive under pressure \cite{Grochala-CSR-2007}. The formation of stable xenon oxides and silicates could explain the missing xenon paradox - the observation that the amount of Xe in the Earth atmosphere is an order of magnitude less than what it would be if all Xe were degassed from the mantle into the atomosphere \cite{Anders-Science-1977}.One explaination for this deficit is that Xe is largely retained in the mantle. In fact, it has been reported that a few weight percent of xenon can be incorporated into SiO$_2$ at elevated pressures and high temperatures \cite{Sanloup-GRL-2002, Sanloup-Science-2005}. A recent theoretical investigation has shown that no xenon carbides are stable up to at least 200 GPa \cite{Oganov-Psi-2007}, and an experimental and theoretical high pressure study \cite{Caldwell-Science-1997} found no tendency for xenon to form a metal alloy with iron or platinum.

Here we investigate the possible stability of xenon oxides using quantum-mechanical calculations of their energetics. As the structures of stable xenon oxides are not experimentally known, we calculate them using the recently developed evolutionary algorithm for crystal structure prediction \cite{Oganov-JCP-2006, Oganov-ACR-2011}. We also analyse chemical bonding in these exotic compounds. 

We have performed structure prediction simulations for the Xe-O system for the compositions of {XeO, XeO$_2$, XeO$_3$, XeO$_4$} at 5, 50, 100, 120, 150, 180, 200 and 220 GPa. Our calculation at 5 GPa yielded lowest-enthalpy structures that always contained the O$_2$ molecules, indicating the tendency for segregation of the elements, and indeed at 5 GPa decomposition was found to be energetically favourable. This suggests that the reaction observed by Sanloup et al. \cite{Sanloup-GRL-2002, Sanloup-Science-2005} at 0.7 to 10 GPa was an entropically driven incorporation of Xe impurities into the structure of SiO$_2$, rather than enthalpically-driven formation of a stoichiometric xenon silicate or oxide. Indeed, solid solutions and point defects are stabilized by entropy \cite{Urusov-1977}.

\begin{figure*} [!htb]
\epsfig{file=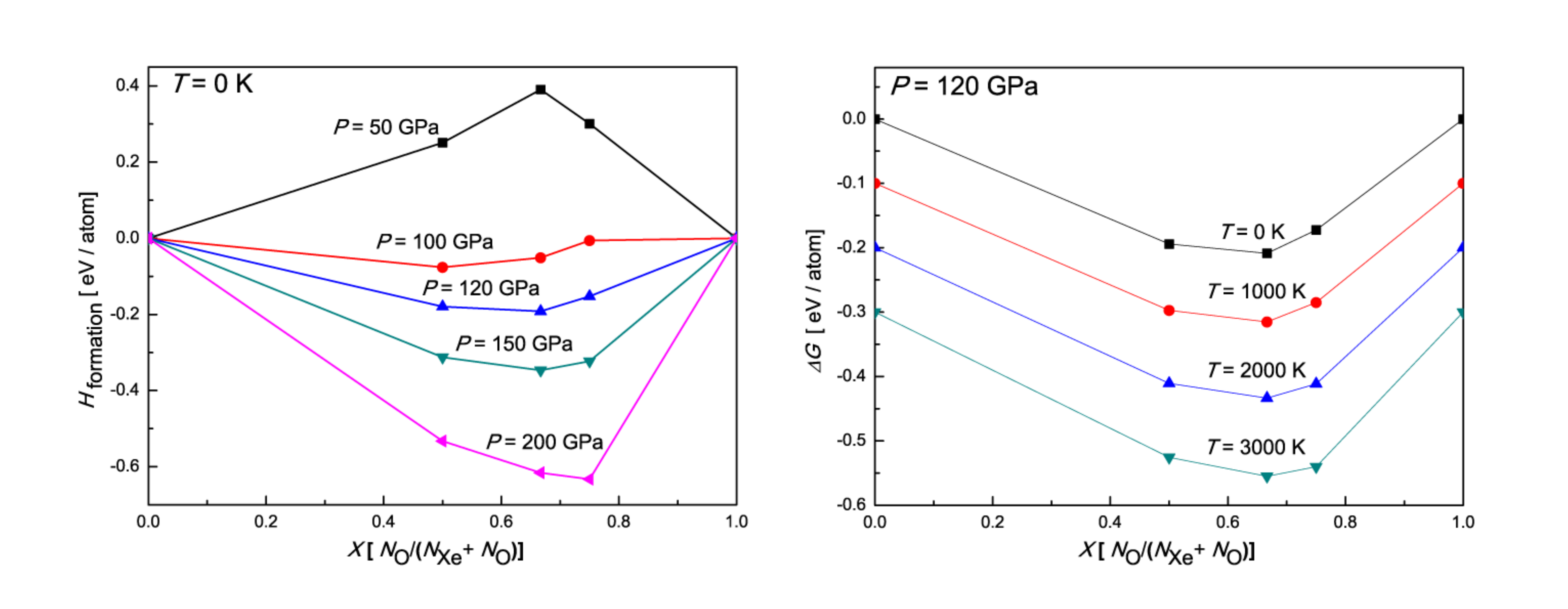, width=0.7\textwidth}
\caption{\label{formation} a) Predicted enthalpy of formation of Xe-O compounds at high \emph{P} and \emph{T}=0 K; b) Predicted Gibbs free energy of formation of Xe-O compounds at different temperatures and \emph{P}=120 GPa. The compounds shown are, from left to right, Xe, XeO, XeO$_2$, XeO$_3$, O. For oxygen, the structures of the $\zeta$-phase \cite{Ma-PRB-2007} and $\epsilon$-phase \cite{Lundegaard-Nature-2006} were used. For Xe, the fcc \cite{Sears-JCP-1962} and hcp \cite{Sonnenblick-CPL-1977} structures were considered and hcp was found energetically more favourable above 100 GPa, in agreement with experiments \cite{Boehler-PRB-1996}. }
\end{figure*}

However, detailed calculations at and above 100 GPa indeed found a number of oxides stable against decomposition into the elements. The most promising structures and compositions were investigated further. The Gibbs free energy of formation of these oxides at high \emph{P, T} conditions is shown in Fig. \ref{formation}. At 0 K, XeO, XeO$_2$ and XeO$_3$ at high \emph{P} become thermodynamically stable. We also performed quasi-harmonic free energy calculations with phonon spectra using the finite-displacement method \cite{phononpy-2008} to assess the effect of temperture on their stability (Fig. \ref{formation}b). The inclusion of thermal effects does not bring any quantitative changes into the picture, and XeO, XeO$_2$ and XeO$_3$ remain thermodynamically stable also at high \emph{T}. 

\begin{figure*}
\epsfig{file=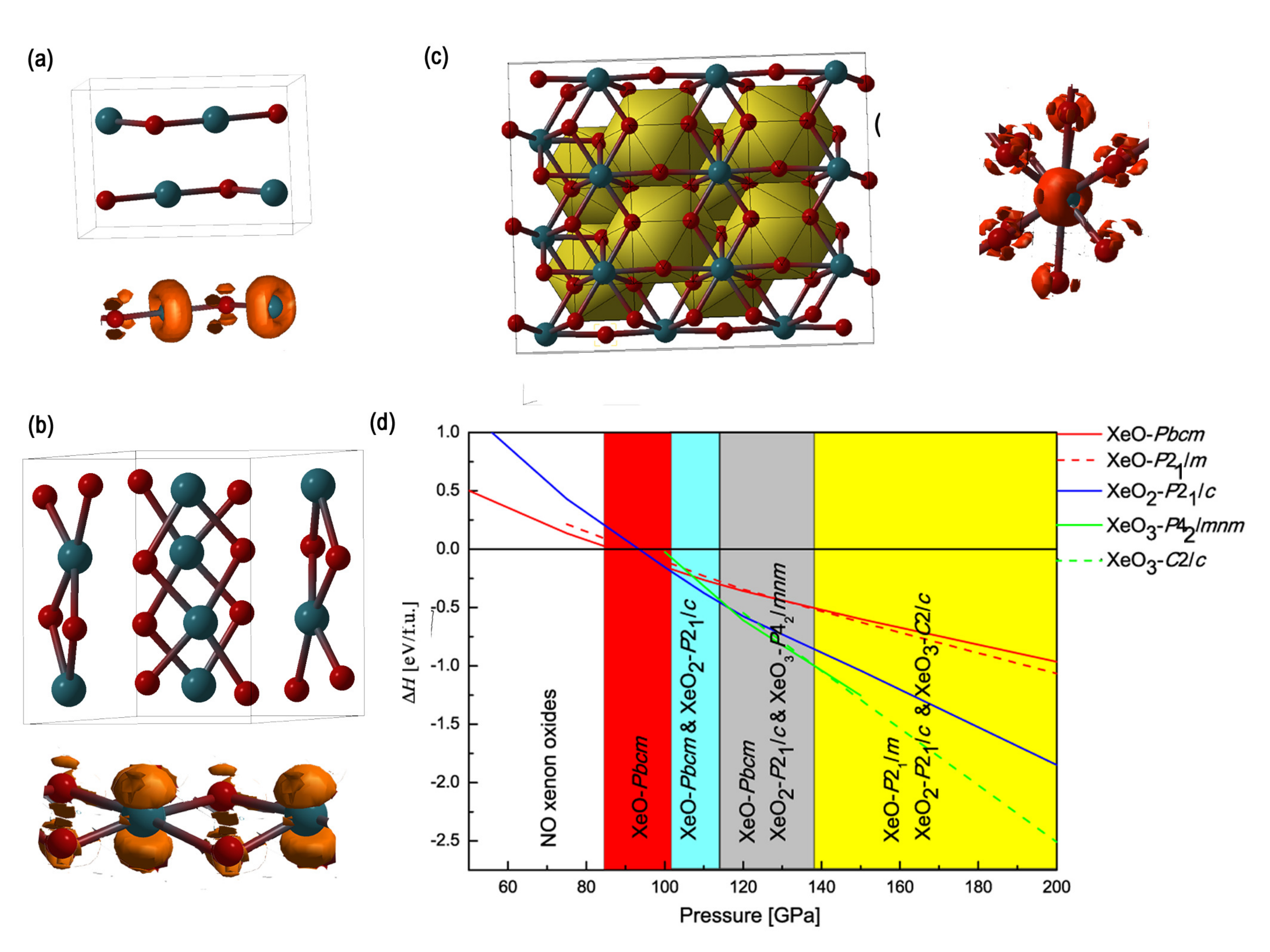, width=0.7\textwidth}
\caption{\label{ELF} a) Crystal structure of XeO (\emph{Pbcm}) Structure at 100 GPa, and its ELF picture ($\rho$ = 0.85) on the Xe-O chain; b) Crystal structure of XeO$_2$ (\emph{P}2$_{1}$/\emph{c}) structure at 120 GPa, and its ELF picture ($\rho$ = 0.85) on the XeO$4$ square. c) crystal structure of  XeO$_3$ (\emph{Pmmn}) Structure at 200 GPa, and its ELF picture ($\rho$ = 0.82) on XeO$_{12}$ anticuboctahedra; d) Enthalpies of formation (solid lines) and enthalpies of decomposition into other xenon oxides (dashed lines) of all stable xenon oxides. Stability fields are delineated by colors (At 198 GPa, \emph{P}2$_{1}$/\emph{c} XeO$_2$ transforms to a \emph{Cmcm} structure, and \emph{C}2/\emph{c}-XeO$_3$ transforms to \emph{Pmmn} phase, i.e., just at the edge of the graph.}
\end{figure*}

Fig. \ref{ELF}d shows the enthalpy of formation of all the Xe oxides as a function of pressure. Below 83 GPa all xenon oxides are unstable. At 83 GPa, XeO-\emph{Pbcm} becomes stable, followed by XeO$_2$-\emph{P}2$_{1}$/\emph{c} above 102 GPa and XeO$_3$-\emph{P}4$_2$/\emph{mnm} above 114 GPa. There is an interesting trend for the oxidation number of Xe to increase with increasing pressure.

A simple and clear analysis of chemical bonding can be done using the electron localization function (ELF) \cite{Becke-JCP-1990}. The ELF gives information about the valence electron configuration of an atom in a compound. States with closed shell electron configurations(Xe$^0$, 5$s^2$5$p^6$, and Xe$^{6+}$, 5$s^2$) will exhibit a spherical ELF distribution, whereas open shell states (Xe$^{2+}$, Xe$^{4+}$) will not. For Xe$^{2+}$ one \emph{p}-orbital is empty and the ELF will have a toroidal shape; likewise, Xe$^{4+}$ can be formed by the removal of two \emph{p}-orbitals and the ELF will show a two-lobe maximum corresponding to the shape of the lone \emph{p}-electron pair.

The most stable structure of XeO at 100 GPa has space group \emph{Pbcm} and 8 atoms in the unit cell. As shown in Fig. \ref{ELF}a, Xe atoms are in a twofold (linear) coordination and Xe-O bonds form chains, with O-Xe-O angles of 175.6$^\circ$ and Xe-O-Xe angles of 112.6$^\circ$. The alternating Xe-O bond lengths are 2.0 and 2.1 \AA. The ELF picture (Fig. \ref{ELF}a) shows a toroidal maximum of ELF around each Xe atom, exactly what one should expect for Xe$^{2+}$ state. Above 145 GPa XeO undergoes a phase transformation and forms a structure with the space group \emph{P}2$_1$/\emph{m}. Xe1 connects to four oxygens and has square coordination, and forms thus the same chains as in XeO$_2$ (suggesting that Xe1 atoms are in the tetravalent Xe$^{4+}$ state), whereas Xe2 can be described as neutral and not bonded to other atoms by any significant bonds. The presence of neutral non-bonded atoms in this structure is energetically favourable as it increases its packing density. 

For XeO$_2$, the stable structure above 102 GPa has space group \emph{P}2$_{1}$/\emph{c} and 24 atoms in the unit cell. Xenon atoms have a slightly non-planar square coordination and the structure consists of 1D-ribbons of edge-sharing XeO$_4$-squares (Xe-O distances are 2.0 and 2.1 \AA), with four Xe-O bonds and two lone-pair maxima forming an octahedron, consistent with the geometry proposed by recent experiment \cite{Brock-JACS-2011}. Just as in XeO, there are no peaks visible in the ELF isosurface along the Xe-O bonds (Fig. \ref{ELF}b). Above 198 GPa XeO$_2$ transforms into the XeO$_2$-\emph{Cmcm} structure. This non-trivial structure can be represented as having parallel zigzag Xe chains (Xe-Xe distances are 2.62 \AA), with each Xe atom having two neighbouring oxygen atoms in the form reminiscent of bent (the Xe-O distance is 1.95 \AA, O-Xe-O angle is 160.7$^\circ$) XeO$_2$ molecules.

XeO$_3$ becomes stable at 114 GPa. Its structure has space group is \emph{P}4$_2$/\emph{mnm} and 16 atoms in the unit cell. It is stable towards decomposition into Xe and O$_2$ as well as into XeO or XeO$_2$ and O$_2$.  As shown in Supplementary Materials, \emph{P}4$_2$/\emph{mnm} phase is composed of two sublattices: square XeO$_2$ chains again, suggesting the Xe$^{4+}$ state and chains made of linear O$_2$ dumbbells. Above 145 GPa, the molecules in the linear -O$_2$-O$_2$- chains are partly dissociated and we observe the -O$_2$-O- chain in the \emph{C}2/\emph{c} phase with 48 atoms per unit cell. Above 198 GPa, the structure transforms to a \emph{Pmmn} phase with 8 atoms per unit cell (Supplementary Materials). In this remarkable structure, the oxygen atoms form anticuboctahedrea in which the Xe atom sits in the center (Fig. \ref{ELF}c). The ELF distribution around Xe atoms in the \emph{Pmmn} phase is spherical around the xenon, which points at the Xe$^{6+}$ valence state with a spherically symmetric 5\emph{s$^2$} valence shell. Again, we observe the tendency of increasing oxidation states under pressure.

Xenon fluorides are stable at ambient conditions, xenon oxides become stable above 83 GPa, and xenon carbides are unstable up to 200 GPa at least \cite{Oganov-Psi-2007}. It thus appears that xenon forms compounds most readily with the most electronegative atoms, and that in turn suggests that ionicity is essential. This is somewhat counterintuitive, given that xenon atom has a very stable closed valence shell and its electronegativity is rather high. The electronegativity difference (1.4 for Xe-F, 0.8 for Xe-O and 0.56 for Xe-C) determines the degree of ionicity at ambient conditions. However, ionicity could be enhanced under pressure. Spontaneous ionization under pressure was recently found even in elemental boron \cite{Oganov-Nature-2009}. 

\begin{figure*}
\epsfig{file=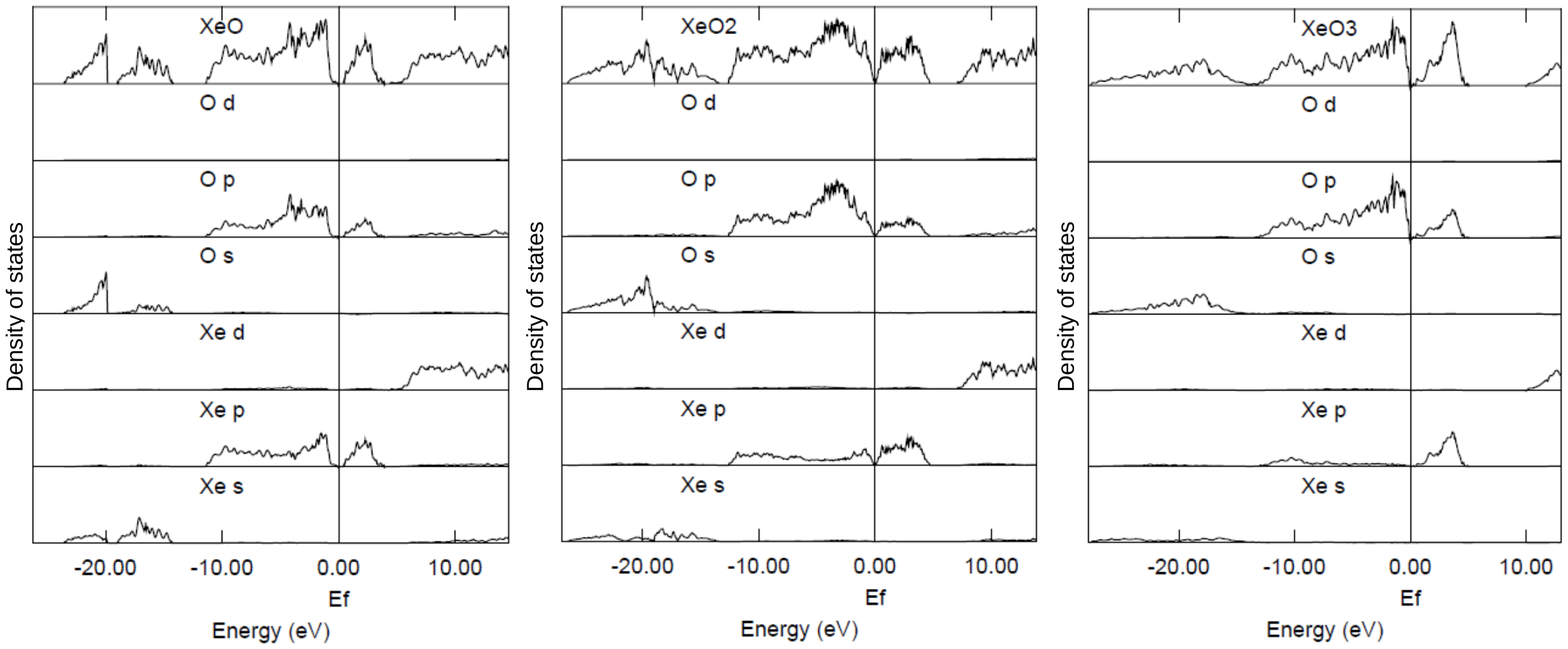, width=0.8\textwidth}
\caption{\label{DOS} Total and projected densities of states for a) XeO-\emph{Pbcm} at 100 GPa, b) XeO$_2$-\emph{P}2$_1$/\emph{c} at 150 GPa, c) XeO$_3$-\emph{Pmmn} at 200 GPa. The sum over the plotted projections gives the total DOS for each system. The DOS plotted have different units on the y-axis so as to give the total number of valence electrons for each system when integrated up to the Fermi level. This corresponds to 56 $e^-$ (XeO, \emph{Z}=4), 160 $e^-$ (XeO$_2$, \emph{Z}=8) and 52 $e^-$ (XeO$_3$, \emph{Z}=2).}
\end{figure*}

To obtain further chemical insight into these exotic xenon oxides, we selected some of the stable structures containing Xe$^{2+}$, Xe$^{4+}$, and Xe$^{6+}$, namely, XeO-\emph{Pbcm} at 100 GPa, XeO$_2$-\emph{P}2$_1$/\emph{c} at 150 GPa, and XeO$_3$ \emph{Pmmn} at 200 GPa. Fig. \ref{DOS} shows the density of states and its projection onto atomic orbitals. All of these xenon oxides are narrow-gap semiconductors. Using state-of-the-art \emph{GW} calculations, we got the band gaps: 1.52 eV for XeO-\emph{Pbcm}, 0.52 eV for XeO$_2$-\emph{P}2$_1$/\emph{c}, 0.15 eV for XeO$_3$-\emph{Pmmn}; these values should be accurate to within 5-10\%.

The highest valence band levels are dominated by \emph{p} orbitals of O and Xe. Both the atom-projected densities of states and energy-decomposed electron densities suggest charge transfer from Xe to O atoms. Note that the contribution of Xe $p$ orbitals strongly decreases with increasing Xe oxidation because of two concurring effects, the change in stoichiometry and the parallel enhanced electron transfer from Xe to O. These two effects are disentangled in the DOS reported in Supplementary Material. The valence states dominated by \emph{p} contributions contain in XeO-\emph{Pbcm} about 40 electrons, with almost equal contributions from O and Xe. The contribution is almost all \emph{p}, namely 4.51 \emph{p} electron per Xe and 4.93 \emph{p} electrons per O atom. In the case of XeO$_2$-\emph{P}2$_1$/\emph{c} there are about 112 electrons, 31.94 from Xe and 80.16 from O, which implies about 4 \emph{e} per Xe and 5 \emph{e} per O, which suggests that about one electron is transferred from a \emph{p}-orbital of Xe into a \emph{p}-orbital of O. Finally, in XeO$_3$ \emph{Pmmn}, there are about 45 electrons in the valence states dominated by \emph{p} orbital contributions, 5.48 \emph{e} from Xe and 30.10 e from O, leading to a further lowering to 2.8 electrons on Xe \emph{p} orbitals and to again about 5 electrons on O \emph{p} orbitals.

The charge transfer issue was also investigated on the basis of the electron density using Bader's analysis \cite{Bader-1990}. For XeO-\emph{Pbcm} the net charge on the Xe atoms is 1.01, while it increases to 1.995 on average in XeO$_2$-\emph{P}2$_1$/\emph{c} and to 2.754 in XeO$_3$-\emph{Pmmn}. Note that the net charge on oxygen atoms stay almost the same in all the three compounds and very close to -1. This corresponds to ionicity of about 50$\%$ in all oxides. Note that the charge transfer described through the observable function, the electron density, and Bader's analysis confirms what was found with using orbital-projected DOS, a totally independent approach.

Bader analysis yields not only the net atomic charges, but also more subtle characteristics such as the three eigenvalues $\mu$$_\emph{i}$ (\emph{i}=1,3) of the traceless quadrupole moment tensor Q($\Omega$), which is a good quantitative indicator of the departure of an atomic basin from sphericity. The eigenvectors associated with the eigenvalues $\mu$$_\emph{i}$ give the principal directions for relative charge accumulation and depletion. For a spherical distribution the $\mu$$_\emph{i}$ are all equal to zero. Deviation from zero indicates asphericity: negative eigenvalues $\mu$$_\emph{i}$ arise from accumulation of charge in the direction associated with the corresponding eigenvector and at the expense of the directions associated with positive $\mu$$_\emph{i}$. Our calculation (Table \ref{Bader}) shows that Xe atoms in XeO$_3$-\emph{Pmmn} are definitely more spherical, their $\mu$$_\emph{i}$ being all close to zero and about one order of magnitude lower than for the Xe atoms in XeO$_2$ and XeO. This agrees with the ELF picture while providing also a quantitative measure of sphericity. XeO-\emph{Pbcm} (with one \emph{p} orbital partially empty) has, in agreement with the orbital picture, one direction of relative charge depletion, associated to $\mu$$_3$, and two directions of unequal relative charge accumulation, associated to $\mu_1$ and $\mu_2$. Finally in XeO$_2$-\emph{P}2$_1$/\emph{c}, there is one direction of relative charge accumulation associated to $\mu_1$ and two of unequal relative charge depletion, associated to $\mu_2$ and $\mu_3$, in agreement with the picture of two \emph{p} orbitals (partially) empty. 

\begin{table*} 
\begin{tabular}{lrrr}
\hline
       System      & $\mu_1$  &  $\mu_2$  &  $\mu_3$ \\
\hline
       XeO         &    -3.49 &     -2.04 &    5.53 \\
       XeO$_2$-Xe1 &    -5.50 &      1.83 &    3.67 \\
       XeO$_2$-Xe2 &    -5.90 &      2.33 &    3.57\\
       XeO$_3$     &    -0.38 &      0.21 &    0.36\\
        \hline
\hline
\end{tabular}
\caption{\label{Bader}Eigenvalues of the traceless quadrupole moment tensor from the atomic Bader intergration. The selected xenon oxide structures are: XeO-\emph{Pbcm} at 100 GPa, XeO$_2$-\emph{P}2$_1$/\emph{c} at 150 GPa, and XeO$_3$-\emph{Pmmn} at 200 GPa }
\end{table*}

How likely is it that xenon oxides (or silicates) could exist in the Earth's lower mantle, thus explaining the missing Xe paradox? As discussed above, Xe oxides are only stable above 83 GPa, i.e. at pressures corresponding to the lower mantle, where metallic Fe should be present \cite{Frost-Nature-2004,Zhang-EPSL-2006}. All of the predicted xenon oxides are very strong oxidants and our calculations show that they will be reduced by Fe, producing iron oxide and free xenon. Then we investigated the formation of xenon silicates under pressure, focusing on XeSiO$_3$ and Xe$_2$SiO$_4$, which contain the least oxidized xenon. All of the investigated compositions were unstable towards decomposition into XeO, XeO$_2$, SiO$_2$ and the elemental Xe; Xe$_2$SiO$_4$ proved to be one of the least unstable silicates, but still is unstable. Our conclusion is: while Xe oxides are stable in the Xe-O system, at strongly reducing conditions neither silicates nor oxides of Xe can exist: Xe silicates are unstable to decomposition, while Xe oxides will be reduced by metallic Fe present in the lower mantle.

To summarize, we have predicted stability of xenon oxides at high pressure, which can be readily tested experimentally. On increasing pressure, increasingly high oxidation states of Xe will appear - first, XeO (above 83 GPa), then XeO$_2$ (above 102 GPa), then XeO$_3$ (above 114 GPa). Present results clearly show that Xe loses its chemical inertness under pressure and that charge transfer plays an essential role in chemical bonding of Xe compounds and their stability is largely determined by electronegativity differences. Furthermore, pressure stabilizes increasing oxidation states of Xe atoms (Xe$^{0}$ $\rightarrow$ Xe$^{2+}$ $\rightarrow$ Xe$^{4+}$ $\rightarrow$ Xe$^{6+}$) and enhances charge transfer from Xe to O atoms. We find that xenon silicates are not stable at pressures of the Earth's mantle ($<$136 GPa) and that xenon oxides, although stable against decomposition into the elements, will be reduced to free Xe in the strongly reducing conditions of the lower mantle. 

Although the formation of stable Xe oxides or silicates is not possible at conditions found in the Earth's mantle, the formation of strong Xe-O bonds under pressure, clearly seen in our results, implies that Xe may still be retained at point or line defects, or grain boundaries of mantle minerals. Xenon could also be stored in perovskite/post-perovskite stacking faults\cite{Oganov-Nature-2005}. The facile chemical bonding between Xe and O atoms demonstrated here and the preference of Xe atoms to terminate the silicate perovskite layers, observed in our simulations (xenon silicate in Supplementary Materials), both suggest this possibility. Indeed, the effect of trapping of trace elements by lattice defects is well known \cite{Urusov-GEO-1998}

\section{Methods}
{\bf Crystal Structure Prediction}: The evolutionary algorithm USPEX \cite{Oganov-JCP-2006, Oganov-ACR-2011}, used here for predicting new stable structures, searches for the structure with the lowest free energy at the given pressure-temperature conditions and is capable of predicting the stable structure of a compound knowing just the chemical composition. Details of the method are described elsewhere \cite{Oganov-JCP-2006, Oganov-ACR-2011} and a number of applications \cite{Oganov-JCP-2006, Oganov-Nature-2009} illustrate its power. All structure prediction runs discussed here were done with the USPEX code, with structure relaxations done using density functional theory (DFT) within the generalized gradient approximation (GGA) \cite{GGA-1996} in the framework of all-electron projector augmented wave (PAW) \cite{PAW-1994} method as implemented in the VASP \cite{VASP-1996} code. We used the plane wave kinetic energy cutoff of 520 eV and the Brillouin zone was sampled with the resolution of 2$\pi$ $\times$ 0.08 \AA$^{-1}$, which showed excellent convergences of the energy differences, stress tensors and structural parameters. We studied systems containing up to 36 atoms per unit cell. The first generation of structures was created randomly. All structures were relaxed at constant pressure and the enthalpy was used as fitness. The energetically worst structures (40$\%$) were discarded and a new generation was created from the remaining structures through heredity, lattice mutation and permutation of atoms. Additionally, the best structure of a generation was carried over into the next generation. We terminated the runs generally after 50 generations, and all runs had found the minimum enthalpy structures much earlier. The population size was set to at least twice the number of atoms in the cell. Results obtained with and without the van der Waals functional \cite{VDW-2011} (as implemented in the VASP code) are very similar, and here we show the results that include this functional.

{\bf Chemical bonding analysis}: The density of states and its projection onto the atomic orbitals have been calculated through the periodic LCAO approach, using the CRYSTAL-06 code \cite{Dovesi-2006} and the same DFT functional \cite{GGA-1996} as used in all calculations described here. Oxygen atoms were described by an 8-411+\emph{d} all electron basis set \cite{Towler-PRB-1994} and Xe atoms by a cc-pVTZ basis set \cite{Peterson-JCP-2003}, specifically devised for the fully-relativistic ECP28MDF pseudopotential, with 26 electrons per Xe kept active. In order to avoid problems of numerical catastrophes \cite{Dovesi-2006} , the original cc-pVTZ(12\emph{s}11\emph{p}9\emph{d}1\emph{f})/[5\emph{s}4\emph{p}3\emph{d}1\emph{f}] basis set \cite{Peterson-JCP-2003} was slightly modified by removing the outermost \emph{s}, \emph{p} and \emph{d} Gaussians and the \emph{f} polarization function to yield a final (11\emph{s}10\emph{p}8\emph{d})/[5\emph{s}4\emph{p}3\emph{d}] contracted basis set. The Kohn-Sham matrix was diagonalized on an isotropic 8 $\times$ 8 $\times$ 8 k-mesh and the same mesh was used in the Fermi energy calculation and density matrix reconstruction. Bader's charges and atomic quadrupole moment tensors were evaluated from the periodic LCAO electron density and using the TOPOND package \cite{Topond-1999} interfaced to the CRYSTAL code. TOPOND determines the boundaries of the atomic basins and the integrated properties within these basins using fully analytical and on-the-fly evaluations of the electron densities and its derivatives (up to the 4th order), i.e. no use is made of electron densities on a grid or of numerical approximations of electron density gradients \cite{Gatti-2007}. Basin boundaries are determined through the PROMEGA algorithm \cite{Keith-1993}, while the basin integration is performed in spherical coordinates, using Gaussian quadrature formulas.  The resulting atomic charges were very close to those obtained from the PAW wavefunction and the numerical algorithm described in \cite{Henkelman-CMS-2006}. Band gaps were recalculated within \emph{GW} approximation as implemented in the VASP code \cite{vasp-GW-2007}.

\begin{acknowledgments}
Calculations were performed on the CFN cluster and Blue Gene supercomputer (Brookhaven National Laboratory), Swiss Supercomputer Centre, Skif MSU supercomputer (Moscow State University) and at the Joint Supercomputer Center of the Russian Academy of Sciences (Moscow). A.R.O. thanks DARPA (grant no. W31P4Q1210008) and the National Science Foundation (grant no. EAR-1114313) for financial support.
\end{acknowledgments}

\bibliographystyle{model1a-num-names}
\bibliography{biblio}
\end{document}